\documentclass[aps,twocolumn,amsmath,amssymb,prl,showpacs]{revtex4}
\usepackage[dvips]{graphicx}
\usepackage{psfrag}
\usepackage{epsfig}

\begin{document}

\title{Pesin-Type Identity for Weak Chaos}
\author{Nickolay Korabel and Eli Barkai}
\affiliation{Physics Department, Bar-Ilan University, Ramat-Gan 52900, Israel}

\date{\today}

\begin{abstract}
Pesin's identity provides a profound connection between entropy $h_{KS}$ (statistical mechanics) and the Lyapunov exponent $\lambda$ (chaos theory). It is well known that many systems exhibit sub-exponential separation of nearby trajectories and then $\lambda=0$. In many cases such systems are non-ergodic and do not obey usual statistical mechanics. Here we investigate the non-ergodic phase of the Pomeau-Manneville map where separation of nearby trajectories follows $\delta x_t= \delta x_0 e^{\lambda_{\alpha} t^{\alpha}}$ with $0<\alpha<1$. The limit distribution of $\lambda_{\alpha}$ is the inverse L{\'e}vy function. The average $\left< \lambda_{\alpha} \right>$ is related to the infinite invariant density, and most importantly to entropy. Our work gives a generalized Pesin's identity valid for systems with an infinite invariant density. 
\end{abstract}

\pacs{05.90.+m, 05.45.Ac, 74.40.+k}

\maketitle

Chaotic systems are characterized by exponential separation of nearby trajectories, which is quantified by a positive Lyapunov exponent $\lambda$ \cite{Dor}. Such a behavior leads to the need for statistical approaches since chaos implies our inability to predict the long time limit of a system in a deterministic fashion. A profound relation between chaos and statistical mechanics is given by Pesin's identity \cite{Dor}. It states that the Kolmogorov-Sinai entropy $h_{KS}$ is equal to the Lyapunov exponent $\lambda$ for a closed ergodic 1d systems (to the sum of positive Lyapunov exponents for $d>1$). At the same time, it is well known that many systems such as Hamiltonian models with mixed phase space \cite{Zaslavsky}, systems with long range forces \cite{Latora}, certain billiards \cite{Li} and one-dimensional hard-particle gas \cite{Grassberger} have a Lyapunov exponent equal zero. While for complex systems it may be extremely difficult to determine whether the Lyapunov exponent is zero or small, due to numerical inaccuracies, it turns out that most fundamental text book examples of chaos theory {\em may} have a zero Lyapunov exponent. Prominent examples for such {\em weakly chaotic} systems are the logistic map at the edge of chaos (Feigenbaum's point) \cite{Tsallis} and the Pomeau-Manneville map which is used to model intermittency (originally in turbulence) \cite{GW88}. If the Lyapunov exponent is zero, i.e.\ separation of trajectories is sub-exponential, we have a strong indication that the usual Boltzmann-Gibbs statistical mechanics is not valid. Indeed it was found that certain systems with zero Lyapunov exponents break ergodicity \cite{BB06,RB07}. Classical entropy theory is also not applicable in this case \cite{GW88, Tsallis}, particularly the entropy and average algorithmic complexity grow non linearly in time \cite{GW88}, while for a system with a positive Lyapunov exponent they increase linearly in time. Still the situation is not hopeless from the point of view of statistical mechanics and one may consider distributions of time average observables \cite{BB06,RB07,Akimoto,TA93,TZ06}. 

Connection between possible generalizations of usual statistical mechanics and weakly chaotic systems have attracted much attention recently. In particular, a generalized Pesin's identity for the logistic map at the edge of chaos was investigated using Tsallis statistics \cite{Tsallis,Robledo}. A critical discussion of this approach is given in Ref.\ \cite{Gra05} (and see a reply in \cite{Rob06}). According to \cite{Gra05} a meaningful generalized Pesin's identity must satisfy certain requirements. (i) Averages must be made with respect to the natural density, in our case the {\em infinite invariant density} \cite{Aaronson,Thaler} (see details below). (ii) Evolution should be for long times, unlike previous attempts to generalize Pesin's identity. (iii) The relevant entropy is the entropy of Kolmogorov and Sinai, not Boltzmann-Gibbs \cite{Vulp}. (iv) Results should be general in that they do not depend on particular initial conditions. The generalized Pesin's identity, Eq.\ (\ref{GPesin}) below, fulfills these requirements. Thus, we establish a profound relation between separation of nearby trajectories and entropy, even though the separation is sub-exponential. 


Consider the Pomeau-Manneville map \cite{PM80} on the unit interval with one marginally unstable fixed point located at $x=0$ 
\begin{equation}
\label{map_eq_1} 
M(x_t) =  x_t + a x_{t}^{z}  \; \left(mod. \; 1\right), \; \; z\ge1, \; a>0.
\end{equation}
The discontinuity point $\xi$ is defined by $M(\xi)=1$. This map is one of the pioneer models of intermittency. Its generalizations attracted vast research using different methods such as continuous time random walks \cite{GT84,Zum93} and periodic orbit theory \cite{Artuso} to name a few. Sojourn times of trajectories in the vicinity of the unstable fixed point are described by power law statistics leading to aging \cite{Barkai03} and non Gaussian fluctuations \cite{GW88}, which are related to weak ergodicity breaking \cite{BB06,RB07}. 

In the non ergodic phase $z>2$ the density function of the map is concentrated on the unstable fixed point in the long time limit. The derivative $\left| M'(x) \right|$ at this point is equal to $1$, so $\lambda=0$ as shown already in \cite{GW88}. Such a behavior is found since most of the time the particle spends in the vicinity of the marginally stable fixed point. Following \cite{GW88} assume that the sensitivity of nearby trajectories is stretched exponential $\delta x_t= \delta x_0 e^{\lambda_{\alpha} t^{\alpha}}$ with $0<\alpha<1$. Using the chain rule, and the dynamical mapping $x_{t+1}=M(x_t)$ we have
\begin{equation}
\label{eq_1} 
\lambda_{\alpha}(x_0) = \frac{1}{t^{\alpha}}\sum_{i=0}^{t-1} \ln \left| M'(x_i) \right|,
\end{equation}
where the dependence on initial condition is emphasized. For normal case and an ergodic system we have $\alpha=1$. Then the usual Lyapunov exponent is \cite{Dor}
\begin{equation}
\label{Lyap}
\lambda = \lim_{t\rightarrow \infty} \frac{1}{t} \sum_{i=0}^{t-1} \ln \left| M'(x_i) \right| = \int dx \; \rho(x) \ln \left| M'(x) \right|,
\end{equation}
where $\rho(x)$ is the invariant density of the system. Ergodicity ensures that the time average is equal to the ensemble average. 
To prove that a system actually exhibits stretched exponential separation of trajectories, it is sufficient to find the limit distribution of $\lambda_{\alpha}(x_0)$ and show that it is not trivial (i.e. $0<\alpha<1$).
As we now show in the non ergodic phase $\lambda_{\alpha}$ does not converge to a constant but remains a random variable. Below we obtain the distribution of $\lambda_{\alpha}$ and for this we now calculate the density of trajectories $\rho(x,t)$ of Eq.\ (\ref{map_eq_1}). 

To obtain the density analytically we use the approximation of the map \cite{Grigo} for $x\ll1$, $dx_t/dt \simeq a x_{t}^{z}$, and extend it to be valid on the interval $(0,\xi)$. When the trajectory reaches the boundary $x=\xi$ it is randomly reinjected back to the interval $(0,\xi)$. The density function $\rho_c(x,t)$ of this system is governed by the equation
\begin{equation}
\label{g_Man}
\frac{\partial \rho_c(x,t)}{\partial t} = - \frac{\partial}{\partial x}\left(a x^z \rho_c(x,t)\right) + a \xi^z \rho_c(\xi,t),
\end{equation}
where the subscript $c$ in $\rho_c(x,t)$ is for continuous approximation. The first term on the LHS represents deterministic escape from the marginally unstable fixed point while the second term accounts for reinjection of particles. The solution of Eq.\ (\ref{g_Man}) in Laplace space is
\begin{equation}
\label{g}
\tilde{\rho}_c(x,s) = \frac{\xi^{-1}\tilde{O}_x(s)}{1-a \xi^{z-1}\tilde{O}_{\xi}(s)}, 
\end{equation}
where
\begin{equation}
\label{O_s}
\tilde{O}_x(s)= b \; (z-1) \left[1- (bs)^{\frac{1}{z-1}}\Gamma\left(\frac{z-2}{z-1},bs\right)\right], 
\end{equation}
and $b= (z-1)^{-1} a^{-1} x^{1-z}$. Statistics of the system is controlled by $\alpha = 1$ for $z < 2$, $\alpha = \frac{1}{z-1}$ for $z \ge 2$. Considering the small $s$ behavior (equivalent to $t\rightarrow \infty$) and transforming the solution into the time domain, we obtain for $0 < \alpha < 1$ ($z>2$)
\begin{equation}
\label{g_t} 
\rho_c(x,t) \sim \begin{cases}  \frac{a^{\alpha-1} x^{-\frac{1}{\alpha}}}{\alpha^{\alpha}} \; \frac{\sin(\pi \alpha)}{\pi} \; t^{\alpha-1}, & x \ne 0 \cr
\frac{\sin(\pi \alpha)}{\pi \alpha^{1+\alpha}}\; t^{\alpha}, & x = 0.
\end{cases}
\end{equation}
For $z<2$ we find the following solution
\begin{equation}
\label{g_t2}
\rho_c(x,t) \sim \begin{cases} (2-z) \; x^{1-z}, & x \ne 0 \cr
(2-z) \; t, & x = 0,
\end{cases}
\end{equation}
and for $z=2$ the solution is given by
\begin{equation}
\label{g_t3}
\rho_c(x,t) \sim \begin{cases} \frac{x^{-1}}{\ln(t)}, & x \ne 0 \cr
\frac{t}{\ln(t)}, & x = 0.
\end{cases}
\end{equation}
Note, that the density function is time independent only for $z<2$ and $x \ne 0$ Eq.\ (\ref{g_t2}). We introduce the {\em infinite invariant density} 
\begin{equation}
\label{g_inv}
\bar{\rho}_c(x) = t^{1-\alpha} \; \rho_c(x,t) = \begin{cases}  \frac{a^{\alpha-1}}{\alpha^{\alpha}} \; \frac{\sin(\pi \alpha)}{\pi} \; x^{-\frac{1}{\alpha}}, & x \ne 0 \cr
\frac{\sin(\pi \alpha)}{\pi \alpha^{1+\alpha}}\; t, & x = 0,
\end{cases}
\end{equation}
for $0 < \alpha < 1$. 
Scaled functions $\bar{\rho}_c(x)$ are independent of time for $x\ne0$. Note, that $\bar{\rho}_c(x)\sim x^{-1/\alpha}$, and its integral diverges, $\int_{0}^{\xi} dx \bar{\rho}_c(x) = \infty$. Thus, $\bar{\rho}_c(x)$ is not normalizable \cite{Aaronson,Thaler}. Still, as we show later, the infinite invariant density is useful for the calculation of the statistical properties of the dynamics. 
\begin{figure}[tbp]
\centerline{\psfig{figure=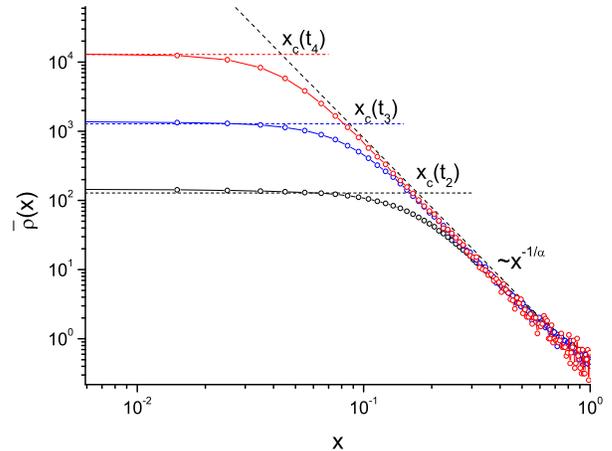,width=90mm,height=70mm}}
\caption{Numerically calculated $\bar{\rho}(x)=t^{1-\alpha} \rho(x,t)$ for the map Eq.\ (\ref{map_eq_1}) with $\alpha=0.3$ and $a=1$. Here $t_i=10^i$, from bottom to top $i=2,3,4$. Dashed lines correspond to Eq.\ (\ref{g_inv}) with no fitting parameters. $x_c$ represents the crossover from one asymptotic to another. Note, that $x_c$ decreases with time, and when $t\rightarrow \infty$ we approach the infinite invariant density.}
\label{Fig_inv_den}
\end{figure}

We compute the invariant density numerically and compare it with the analytical density function Eq.\ (\ref{g_inv}). In these simulations we start with a uniform density and plot $\bar{\rho}(x)=t^{1-\alpha}\rho(x,t)$ versus $x$. Results are shown in Fig.\ \ref{Fig_inv_den}. We find excellent agreement between Eq.\ (\ref{g_inv}) and numerics without fitting. Horizontal lines represent asymptotic solution for $x=0$ calculated for the corresponding time of the simulation, while the sloping line corresponds to the asymptotic solution for $x \ne 0$ which decays as $x^{-\frac{1}{\alpha}}$ (see Eq.\ (\ref{g_inv})). In Fig.\ \ref{Fig_inv_den} $x_c$ represents the crossover from one asymptotic of $\bar{\rho}(x)$ to another. As $t \rightarrow \infty$, $x_c \rightarrow 0$ and we approach the infinite invariant density.
 
\begin{figure}[tbp]
\centerline{\psfig{figure=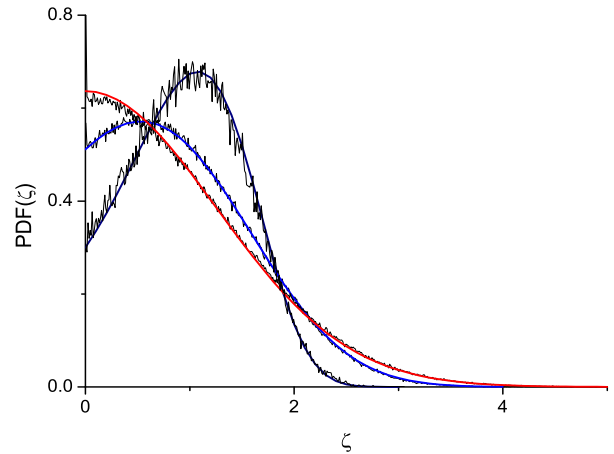,width=90mm,height=70mm}}
\caption{$P_{\alpha}(\zeta)$ for $\alpha = 0.75,0.59,0.5$ from bottom to top on the LHS of the figure obtained with $\left< \lambda_{\alpha} \right>$ calculated according to Eq.\ (\ref{L_mean_4}). Here $t = 10^5$. Smooth curves correspond to analytical PDF Eq.\ (\ref{renewal_pdf_zeta}) without fitting.}
\label{Manneville_PDF_L}
\end{figure}

  
To find the distribution of $\lambda_{\alpha}(x_0)$ Eq.\ (\ref{eq_1}), we use a simple stochastic approach. The same distribution can be found using Darling-Kac theorem applied to our observable Eq.\ (\ref{eq_1}) \cite{TZ06}. Consider the logarithm of the derivative of the map $y=\ln \left| M'(x_t) \right|$ in the non ergodic phase $0<\alpha<1$. We define a two state process $I(t)=0$ if $y<\xi$ and $I(t)=1$ if $y>\xi$. Waiting times in state $0$ are distributed according to the probability density function (PDF) $\psi(t)\sim A/t^{1+\alpha}$ as $t \rightarrow \infty$, or in Laplace space $\tilde{\psi}(s)=\int_0^{\infty}dt e^{-st} \psi(t)\sim 1 - B s^{\alpha}$, as $s \rightarrow 0$, where $A,B$ are positive constants, so the average waiting time is infinite as well known \cite{GT84,Zum93}. In contrast, waiting times in state $1$ have a characteristic average time. Neglecting correlations, we consider $I(t)$ as a renewal process. Let $n$ be the number of renewals, namely, number of transitions from state $0$ to $1$. The logarithm of the derivative of the map $\ln \left| M'(x_t) \right|$ is equal to zero most of the time (roughly when $I(t)=0$) since the trajectory stays for long time near marginally unstable fixed points, only for short periods its value deviates from zero. The sum of logarithms along a trajectory is thus proportional to $n$: $\sum_{i=0}^{t-1} \ln \left| M'(x_i) \right| \sim c \; n$, where $c$ is a positive constant. Our goal is to calculate the PDF of scaled generalized Lyapunov exponents $\lambda_{\alpha}$ Eq.\ (\ref{eq_1}). The PDF of the number of renewals $n$ which occur up to time $t$ is given by \cite{Feller}
\begin{equation}
\label{renewal_pdf}
P_n(t) = \frac{1}{\alpha} \frac{t}{n^{1+1/\alpha}B^{1/\alpha}} l_{\alpha}\left[\frac{t}{\left( Bn\right)^{1/\alpha}}\right],
\end{equation}
where $l_{\alpha}$ is the one-sided L{\'e}vy PDF defined through its Laplace transform $\tilde{l}_{\alpha}(s)= exp(-s^{\alpha})$. We define $\zeta=\lambda_{\alpha}/\left< \lambda_{\alpha} \right>$ and, since $\left< \lambda_{\alpha} \right> = c \left< n \right>$, $\zeta=n/\left< n \right>$ is independent of $c$. Using Eq.\ (\ref{renewal_pdf})
\begin{equation}
\label{renewal_pdf_zeta}
P_{\alpha}(\zeta) = \frac{\Gamma^{1/\alpha}(1+\alpha)}{\alpha \zeta^{1+1/\alpha}} \; l_{\alpha}\left[\frac{\Gamma^{1/\alpha}(1+\alpha)}{\zeta^{1/\alpha}}\right].
\end{equation}
This is one of the main equations in the manuscript since it gives the PDF of scaled generalized Lyapunov exponents $\lambda_{\alpha}/\left<\lambda_{\alpha}\right>$. Distributions of $\zeta=\lambda_{\alpha}/\left< \lambda_{\alpha} \right>$ obtained by simulations are shown in Fig. \ref{Manneville_PDF_L}. Smooth curves correspond to analytical PDF Eq.\ (\ref{renewal_pdf_zeta}) without fitting. The perfect agreement between theory and numerical results indicates that the general theory works well for finite time simulations.

Now we calculate the average $\left< \lambda_{\alpha} \right>$. Using Eq.\ (\ref{eq_1})
\begin{equation}
\label{L_mean}
\left< \lambda_{\alpha} \right> = \int_0^1 \frac{\sum_{i=0}^{t-1} \ln \left| M'(x_i) \right|}{t^{\alpha}} \; \rho(x_0) \; dx_0,
\end{equation}
where the averaging is over initial conditions distributed according to some initial density. Since we are interested in the long time limit, we replace the summation with an integral and average over the density function 
\begin{equation}
\label{L_mean_1}
\left< \lambda_{\alpha} \right> \sim \frac{1}{t^{\alpha}} \int_0^1 dx \int_{0}^{t} \ln \left| M'(x) \right|\; \rho(x,t) \; dt.
\end{equation}
According to Eq.\ (\ref{g_t}), the density function has two asymptotics valid for $x=0$ and $x\ne0$ 
\[
\left< \lambda_{\alpha} \right> \sim \frac{1}{t^{\alpha}} \int_0^{x_c} dx \int_{0}^{t} \ln \left| M'(x) \right|\; \rho(0,t) \; dt +
\]
\begin{equation}
\label{L_mean_2}
+ \frac{1}{t^{\alpha}} \int_{x_c}^1 dx \int_{0}^{t} \ln \left| M'(x) \right|\; \rho(x,t) \; dt,
\end{equation}
where $x_c$ denotes the crossover from one asymptotic of $\rho(x,t)$ to another (see Fig.\ \ref{Fig_inv_den}).\ We define it as $\rho(x_c)=\rho(x=0)$. Using Eq.\ (\ref{g_t}), $x_c = \alpha^{\alpha} t^{-\alpha}$.
When $t \rightarrow \infty$, $x_c \rightarrow 0$, so the first integral in Eq.\ (\ref{L_mean_2}) vanishes. This fact is not obvious since $\rho(x=0) \rightarrow \infty$ (see Eq.\ (\ref{g_t})), it happens because we consider a specific observable with $\ln \left| M'(x) \right| \rightarrow 0$ as $x\rightarrow 0$, which cancel the $t^{\alpha}$ divergence found in Eq.\ (\ref{g_t}). Computing the integral over time and using $t^{1-\alpha}\rho(x,t)=\bar{\rho}(x)$, we obtain our second main result
\begin{equation}
\label{L_mean_4}
\left< \lambda_{\alpha} \right> = \frac{1}{\alpha} \int_0^1 dx \ln \left| M'(x) \right|\; \bar{\rho}(x).
\end{equation}
Thus, even though $\bar{\rho}(x)$ is not normalizable it yields the average generalized Lyapunov exponent $\left< \lambda_{\alpha} \right>$. Since $\ln \left| M'(x) \right|$ vanishes precisely where the infinite invariant density diverges, the integral is finite and positive. Our main result Eq.\ (\ref{L_mean_4}) is very elegant, since up to a constant $\alpha$ it states that all one needs to describe separation of trajectories is to replace the invariant density with the infinite invariant density. For the stochastic model Eq.\ (\ref{g_Man}) with $0<\alpha<1$ using Eq.\ (\ref{g_inv}), we find
\begin{equation}
\label{L_mean_c}
\left< \lambda_{\alpha} \right> = \frac{1}{\alpha} \int_0^1 dx \; \frac{a^{\alpha-1}}{\alpha^{\alpha}} \frac{\sin(\pi \alpha)}{\pi} \; \frac{\ln (1+ a z x^{\frac{1}{\alpha}})}{x^{1/\alpha}} .
\end{equation}
We emphasize that our main results Eqs.\ (\ref{renewal_pdf_zeta}, \ref{L_mean_4}) are generally valid for systems with an infinite invariant density $\bar{\rho}(x)$. Eq.\ (\ref{L_mean_c}) is specific to maps with one unstable fixed point and can be used to verify the theory numerically.  

\begin{figure}[t]
\centerline{\psfig{figure=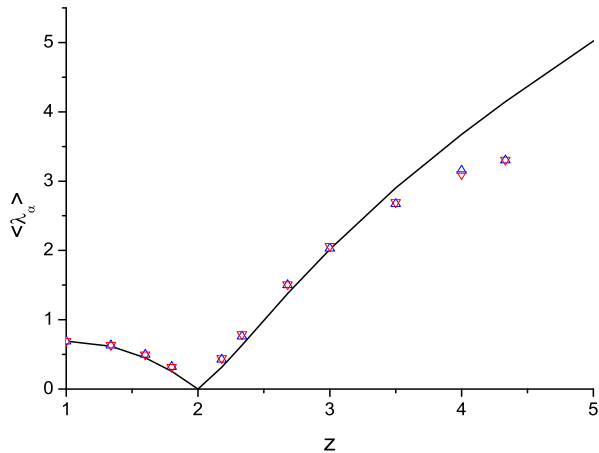,width=90mm,height=70mm}}
\caption{Numerical average of $\left< \lambda_{\alpha} \right>$ (triangles up) perfectly agree with $\left< \lambda_{\alpha} \right>$ (triangles down) calculated by Eq.\ (\ref{L_mean_4}) when $\bar{\rho}$ is calculated numerically. Here $t=10^5$. Solid line is $\left< \lambda_{\alpha} \right>$ Eq.\ (\ref{L_mean_c}) found with approximate infinite invariant density function.}
\label{Manneville_GLE}
\end{figure}

In Fig.\ \ref{Manneville_GLE} we present perfect  agreement between numerical simulations of $\langle \lambda_\alpha \rangle$, and Eq.\ (\ref{L_mean_4}) with $\bar{\rho}(x)$ calculated numerically. For not too large $z$, good agreement between simulations and the theory based on the stochastic  approximation for the infinite invariant density Eq.\ (\ref{L_mean_c}) is found. For large $z$ the convergence is slowed down, since $\alpha$ is small. In the ergodic phase $z<2$ the standard Lyapunov exponent Eq.\ (\ref{Lyap}) is recovered. 


For weakly chaotic systems, $\left< \lambda_{\alpha} \right>$ can be used to extend Pesin's identity. Mathematicians have rigorously shown that entropy for the maps with infinite invariant measure $h_{\alpha}$ satisfies Rohlin's formula \cite{Rohlin,Thaler,Zwei,Zwei00}
\begin{equation}
\label{h_Rohlin}
h_{\alpha}= \int dx \; \bar{\rho}(x) \ln \left| M'(x) \right|, 
\end{equation}
where $\bar{\rho}(x)$ is the infinite invariant density. From Eqs.\ (\ref{L_mean_4}, \ref{h_Rohlin}) we obtain the identity
\begin{equation}
\label{GPesin}
h_{\alpha} = \alpha \left< \lambda_{\alpha} \right>.
\end{equation}
Thus, we have established a profound relation between statistical mechanics and chaos theory for weakly chaotic systems in terms of Pesin-type identity Eq.\ (\ref{GPesin}). The connection between $\left<\lambda_{\alpha}\right>$ and $h_{\alpha}$ is in fact deeper than Eq.\ (\ref{GPesin}), because according to Eq.\ (\ref{renewal_pdf_zeta}) the distribution of the generalized Lyapunov exponents $\lambda_{\alpha}$ is controlled by the scale $h_{\alpha}$ since $\zeta=\lambda_{\alpha}/\left<\lambda_{\alpha}\right>=\alpha \lambda_{\alpha}/h_{\alpha}$. Only in the limit $\alpha \rightarrow 1$ from Eq.\ (\ref{renewal_pdf_zeta}) we get $\lim_{\alpha \rightarrow 1}P_{\alpha}(\lambda)=\delta(\lambda-h_{KS})$. 

Our approach opens up many possibilities of finding connections between dynamical systems characterized by weak chaos and weak ergodicity breaking. Investigation of the relation between weak chaos and transport, generalizing the escape rate formalism \cite{Dor} is one example. Similar to the generalized Lyapunov exponent $\lambda_{\alpha}$ found here, we expect that time averaged transport and diffusion coefficients remain random variables as was recently shown in stochastic models of weak ergodicity breaking \cite{He}. Randomness of $\lambda_{\alpha}$ indicates the modification of the whole structure of statistical mechanics for weakly chaotic systems.

This work was supported by the Israel Science Foundation. We thank R.~Klages, P.~Howard, A.~Robledo and R.~Zweim\"uller for discussions. 

\end{document}